# Hidden dormant phase mediating the glass transition in disordered matter

Eunyoung Park,[1,2,3†] Sinwoo Kim,[1,2,3†] Melody M. Wang,[4] Junha Hwang,[1,2,3] Sung Yun Lee,[1,2,3] Jaeyong Shin,[1,2,3] Seung-Phil Heo,[1,2,3] Jungchan Choi,[1,2,3] Heemin Lee,[1,2,3] Dogeun Jang,[5] Minseok Kim,[5] Kyung Sook Kim,[2,5] Sangsoo Kim,[5] Intae Eom,[2,5] Daewoong Nam,[2,5] X. Wendy Gu,[6] and Changyong Song[1,2,3*]

[1]*Department of Physics, POSTECH, Pohang 37673, Korea*

[2]*Photon Science Center, POSTECH, Pohang 37673, Korea*

[3]*Center for Ultrafast Science on Quantum Matter, Max Planck POSTECH/Korea Research Initiative, Pohang 37673, Korea*

[4]*Department of Materials Science and Engineering, Stanford University, Stanford, California 94305, United States*

[5]*Pohang Accelerator Laboratory, POSTECH, Pohang 37673, Korea*

[6]*Department of Mechanical Engineering, Stanford University, Stanford, California 94305, United States*

[†]*These authors contributed equally to this work.*

[*]*Corresponding author: cysong@postech.ac.kr*

**Metallic glass is a frozen liquid with structural disorder that retains degenerate free energy without spontaneous symmetry breaking to become a solid. For over half a century, this puzzling structure has raised fundamental questions about how structural disorder impacts glass–liquid phase transition kinetics, which remain elusive without**

**direct evidence. In this study, through single-pulse, time-resolved imaging using X-ray free-electron lasers, we visualized the glass-to-liquid transition, revealing a previously hidden dormant phase that does not involve any macroscopic volume change within the crossover regime between the two phases. Although macroscopically inactive, nanoscale redistribution occurs, forming channeled low-density bands within this dormant phase that drives the glass transition. By providing direct microscopic evidence, this work presents a new perspective on the phase transition process in disordered materials, which can be extended to various liquid and solid phases in other complex systems.**

Metallic glasses, a class of vitrified solids lacking long-range translational symmetry, remain in a highly degenerate nonequilibrium state with a complex and poorly understood lattice free energy landscape[1-3]. Their extraordinary structural characteristics have long stimulated intense research into their notoriously enigmatic structure formation process[4-6]. The energetics of structural disorder in metallic glasses share commonalities with other disordered systems such as spin glasses, folded proteins, and polymers, positioning glass as a paragon of disordered materials[7,8]. In addition to these eccentric structural properties, reported high material strength and high elastic limit, particularly at the nanoscale, have enabled a wide range of applications, including precision engineering and nano- and bio-interface components[9,10,11,12].

Large changes, by an order of magnitude of ten or more, in viscosity, dynamic relaxation time, and other properties during the glass transition—distinct from a well-defined thermodynamic transition—have sparked interest in the hidden nature of structure-forming kinetics, challenging our understanding of phase transitions[13-15]. Extensive research has sought to understand disorder as a key feature in describing these puzzling characteristics through multiscale material characterizations, theoretical investigations with new models, and the synthesis of new materials[16-18]. Despite these accumulated discoveries, direct experimental evidence of the structure-forming kinetic process has remained elusive, leaving our understanding speculative. In this study, through ultrafast dynamic X-ray imaging of metallic glass (MG) nanoparticles (NPs) using femtosecond (fs) single pulses from an X-ray free electron laser (XFEL), we directly visualized the glass-to-liquid transition process induced by single fs infrared (IR) laser pulses. The photoexcited hot electrons induced local ionic pressure differences along the laser polarization direction, which evolved to form localized low-density bands that drive the melting of the entire MG NPs. The absence of any macroscopic change has kept this low-density band formation process unnoticed until now. Our discovery of this dormant phase, with solid microscopic evidence, provides new insights into solving the puzzling kinetics that drive such disordered structures.

We conducted ultrafast X-ray imaging experiments to investigate the photoinduced glass transition in MG NPs. The time-resolved investigation was conducted through a pump-probe scheme, using fs-IR lasers as the pump source and fs X-ray pulses for probing[19]. The glass-to-liquid transition of MG NPs was initiated by single fs-IR laser pulses (Ti laser with 800 nm wavelength, 40 fs pulse width, and 400 mJ cm$^{-2}$ fluence, and linearly polarized along the horizontal direction), and the resulting photoinduced structural deformation kinetics were tracked using single-pulse X-ray diffraction imaging (Fig. 1a)[20]. We studied spherical $Ni_xB_{1-x}$ MG NPs with a diameter of 110 nm, loaded onto thin silicon nitride (SiRN) membranes. The spherical MG NPs displayed monodisperse and homogeneous morphology (Fig. 1a, Methods, Fig. S1, and Supplementary Information). Time-resolved single-pulse X-ray diffraction patterns were collected at the Pohang Accelerator Laboratory-XFEL (PAL-XFEL) (Methods).

Upon fs-IR laser illumination, single-pulse X-ray diffraction patterns exhibited anisotropic deformations from the circular Airy ring (Fig. 1b and Fig. S2). The period of fringe oscillations decreased more significantly along the laser field direction compared to the direction perpendicular to the laser field. This anisotropic distortion became more pronounced, displaying smeared fringes along the laser polarization direction at 20 ps. Nanoscale images of the specimens, as plane-projected electron density maps, were obtained by numerically retrieving the phase information from the measured single-pulse X-ray diffraction patterns, clearly visualizing the photoinduced glass-to-liquid transition process, which involved inhomogeneous and dynamic fluctuations of the internal electron density (Fig. 1c, Methods, and Supplementary Video 1)[21]. Rapid redistribution of internal density began at 1 ps (more apparently at 3 ps), with local regions of higher and lower density forming parallel to the laser polarization direction, in contrast to the homogeneous density distribution observed in the intact structure at 0 ps (Fig. S3). This anisotropic density is attributed to localized high-energy spots formed parallel to the laser field, driven by photoexcited surface plasmons[19]. The contrast between the low- and high-density regions became more pronounced by 5 ps as photoinduced thermalization progressed. By 6 ps, significant changes had occurred as the localized low-density regions merged, forming extended channels (black arrows in Fig. 1c). The elongation of these channels through merging continued until 14 ps. From 16 to 20 ps, the channeled low-density regions began to expand laterally, causing an overall expansion of the NP size and leading to eventual disintegration.

As the images clearly show, the transition was accompanied by inhomogeneous density changes, characterized by three distinct development stages. The existence of these stages is demonstrated by the radial expansion of MG NPs estimated from the collected single-pulse images at each delay time (Fig. 2a and Supplementary Information). The data revealed three different transition stages, with a non-expanding, dormant period (stage II) from 6 to 14 ps, separating the other two stages (I and III), which exhibited different expansion rates. Despite minor differences between the perpendicular and parallel directions, the radial expansion rate in stage III (~4 km s$^{-1}$) was more than three times higher than that of stage I (~1.2 km s$^{-1}$). Direct comparison with the thermal expansion of MG materials suggests that the MG NPs were in the glass phase during stage-I and in the liquid state during stage III[22]. Stage II corresponds to the crossover regime in thermal expansion, where the glass transition temperature is typically determined. The observed macroscopic characteristics clearly distinguish the phases, with the dormant period being the one during which the glass-to-liquid transition is induced. The nanoscale images provide microscopic details of the transition process, particularly with the discovery of the new interval (stage II) when the glass transition is initiated.

As previously mentioned, photoexcited MG NPs expand more significantly and at a faster rate along the direction parallel to the laser polarization compared to the perpendicular direction. This asymmetric deformation (by 6% in stage I and further increase to 14% in stage III) is clearly noted in the aspect ratio calculated for the NP diameter parallel ($D_{\parallel}$) to the laser polarization relative to the perpendicular direction ($D_{\perp}$) (Fig. 2b). Notably, the radial expansion in stage I reveals that the expansion along the parallel and perpendicular directions exhibits much different behavior, providing crucial evidence of the characteristic nature of MG materials, which undergo non-affine structural deformation induced by laser-field strains[23]. In contrast to the linear, radial increase observed along the perpendicular direction, the expansion along the laser polarization direction developed parabolically, deviating from the linear increase (Fig. 2c).

This parabolic deviation reflects the plastic deformation in MG caused by localized shear bands, which exhibit serrated flow behavior between stress and strain, a phenomenon commonly observed in MG materials[24]. As reported before, the photoexcited surface plasmon generates a focused, high-energy spot with local pressure accumulation[19,25,26]. It triggers the density redistribution of the MG NP into compressed high-density and expanded low-density regions to induce a dipolar-type shear forces parallel to the laser polarization[27]. Further extension of the low-density regions occurs by relaxing this transiently formed force field, displaying a power-law pattern resulting from the evolution of local shear transformation zones (STZs) into shear bands with a power-law distribution[24]. As the photoinduced deformation proceeds, dynamic heterogeneity decreases due to the cooperative motion of deformation units, which slows down the expansion along the horizontal direction[28]. In contrast, the absence of organized strains and cooperative motion along the perpendicular direction results in linear expansion (Supplementary Information).

To understand the dynamic evolution of heterogeneous density distribution, we calculated the density difference by subtracting the projected density of an ideal sphere of the same size from the obtained images (Fig. 3a, Fig. S7 and Supplementary Information). Regions of relatively lower density (blue) and higher density (red) are clearly manifested in this difference map. The fs-IR laser excitation induced low- and high-density regions immediately from 1 ps, which evolved by enhancing the density contrast and increasing the size of these regions. The spatial alignment of the low- and high-density regions forming a dipole-type shear forces parallel to the laser polarization direction was clearly observed. Upon entering stage II, a new reaction was observed: instead of simple volume expansion, the low-density areas began to stretch out, connecting nearby localized low-density regions to form channeled low-density bands (broken lines in Fig. 3a). This process continued until 14 ps. No particle expansion was observed in stage II, rendering it macroscopically dormant. Instead, the density contrast increased, accompanied by the dynamic rearrangement of low- and high-density regions to form low-density bands. From 16 ps to 20 ps, corresponding to stage III, the low-density bands expanded laterally, perpendicular to the band forming direction (small arrows in Fig. 3a), gradually increasing the overall low-density area and the overall size of the NPs. As the NPs expanded rapidly along with the low-density bands, melting and disintegration of the MG NPs occurred during stage III.

We have quantified the total area of the low-density ($A_{low}$) region, which increased gradually during stage I (Fig. 3b). This areal expansion of the low-density region did not occur during stage II; instead, density redistribution occurred, leading to the formation of connected bands. The non-varying total area of low-density regions in stage II followed the same dormant behavior as observed in the radial expansion (Fig. 2a). Together with the density redistribution (Fig. 1c), this indicates that nanoscale dynamic rearrangement of inhomogeneous density distributions occurs during the glass-to-liquid transition. Importantly, all these active density changes remain hidden macroscopically. This newly discovered dormant phase in stage II, as an intermediate period between the glass and liquid phases, is a unique characteristic of MG not observed during the melting transition of ordinary metals such as gold (Au), where only a monotonic increase in the low-density area is observed (inset in Fig. 3b).

We also estimated the ratio of the low-density area to the total area ($A_{low}/A_{total}$). This ratio exhibits the distinct behavior that the increasing low-density area is not merely a byproduct of particle expansion (Fig. 3c). In stage I of the glass phase, the portion of low-density area, which was more than 50%, increased at a slightly faster rate than the rate of thermalized NP expansion. As stage II approached, this ratio decreased abruptly to 45% and remained constant throughout stage II. NP expansion resumed in stage III (Fig. 2a), but the expansion of the low-density area outpaced the expansion of the NPs, further facilitated by the lateral growth of the low-density band, raising the overall occupancy to the 50% level (Fig. 3c). The melting and disintegration of the NPs in stage III were driven by the extension of the low-density band formed in stage II, acting as a key element in inducing the melting of the MG. This behavior contrasts with the overall monotonic decrease observed in metallic Au, where NP expansion is primarily governed by thermal expansion of the ionic lattice (inset of Fig. 3c)[19].

Notably, the comparison between the changes in the low-density area and the fraction of low-density area with those observed in typical metals like Au confirmed that the three-stage microscopic changes are unique to the phase transition in MG materials.

The overall kinetic process of the photoinduced glass transition is illustrated by focusing on the behavior of the low-density bands (Fig. 4a). The glass-to-liquid transition involves three distinct developmental stages. Hot electrons, photoexcited by fs-IR laser pulses, disrupt the homogeneous density in the intact NPs, inducing localized low- and high-density regions aligned parallel to the IR laser field. This alignment leads to the formation of dipolar shear forces parallel to the IR laser field (gray arrow in Fig. 4a). During stage I of the glass phase, thermalized radial expansion of the NP occurs together with the expansion of low-density regions. Macroscopic particle expansion halts in stage II, which is characterized as a macroscopically dormant state. In this state, active redistribution of the scattered low-density regions formed in stage I occurs, connecting localized regions to form channeled low-density bands. Following the formation of these bands, macroscopic volume expansion resumes at a much higher rate in stage III, corresponding to the liquid phase. This rapid volume expansion in stage III is accompanied by lateral expansion of the low-density bands formed in stage II, ultimately leading to the melting and disintegration of the NPs.

The most notable observation is the discovery of the dormant phase that separates the glass and liquid phases, characterized by active microscopic density redistribution to form low-density bands. This finding indicates that the glass transition point, previously determined somewhat ambiguously as a crossover between the thermal expansion rates of the glass and liquid phases, is not a simple phase transition. Instead, it involves a macroscopically dormant period with active nanoscale density rearrangements that form the bands responsible for melting and disintegration. We further discovered that the channeling of low-density bands is directional, which is explicitly manifested by skeletonizing the bands (Supplementary

Information). A density difference map at 10 ps, overlaid with the skeletons in thick black lines, is shown (Fig. 4b and Fig. S10). The entire skeleton population was quantified for orientation and length, revealing a dominant distribution of the bands between 20 and 40 degrees relative to the laser polarization direction (Fig. 4c). This directional tendency aligns with the shear band in MG materials[29,30,31]. The channeling of low-density bands leads to scattered high-density regions throughout the NPs, which eventually melt and disintegrate. This observation aligns with previous findings, where shear bands in MGs have been shown to consist of inclined low-density regions surrounded by a relatively high-density matrix[30]. This consistency in morphology and angular developments supports the interpretation that the low-density bands observed in our study correspond to the shear bands, further reinforcing the similarities between our results and the well-documented characteristics of shear band formation in MGs[29,32].

The directionality of the low-density bands, or shear bands, is explained through Monte Carlo (MC) simulations. Specifically, the simulations aimed to verify whether the low- and high-density regions, which initially align horizontally along the laser field in a dipole-like configuration, would reorganize into diagonally aligned low-density bands. During this process, the low-density regions preferentially expand along the horizontal direction, corresponding to the direction of the shear force field, compressing neighboring areas and forming high-density regions. As additional density accumulates in the existing high-density regions, they become denser and larger. As the low-density regions continue to propagate, the high-density regions become increasingly compact, making it more challenging for the low-density bands to propagate through or bypass these high-density areas[33,34]. The MC simulation corroborated the experimental observation that the low-density regions do indeed form bands that reorient diagonally relative to the laser polarization (Fig. 4d and Supplementary Information).

Through this study, we provided nanoscale microscopic evidence of the glass-to-liquid phase transition in MG materials using ultrafast X-ray imaging. Of particular interest is the discovery of a dormant period separating the glass and liquid phases, which has been completely unnoticed in previous investigations. During this hidden dormant period, although macroscopically unchanged, active rearrangement of heterogeneous density distribution occurs, forming low-density bands that drive the melting and disintegration of the MG. This process is uniquely observed in MG and is distinctly different from the melting transition seen in metallic crystals such as Au. Unlike crystalline materials, which contain dislocations and undergo strain hardening, MGs localize strain in specific regions where shear bands serve as the primary carriers of plasticity[28]. These localized strains, along with shear bands, deform the material and ultimately lead to the failure of MG. Understanding the formation process of low-density shear bands is thus crucial for preventing their uncontrolled growth and for improving their ductility and fracture resistance[35]. Previous experimental approaches using ensemble-averaged static measurements often failed to capture local heterogeneous dynamics, leaving the nature of the glass transition unclear. To unravel the puzzling structural characteristics of MG materials, it is essential to understand how shear bands evolve in real-time, including in the early stages of low-density region formation that precede macroscopic failure. Our findings address this gap by providing solid microscopic evidence of how nanoscale structural changes drive the glass-to-liquid transition using time-resolved single-particle, single-pulse XFEL imaging. This approach allows us to capture microscopic changes in low-density regions and link them to the material's overall macroscopic behavior. The newly discovered dormant phase represents a critical period in which dynamic density rearrangements, including the formation of low-density shear bands, actively occur as an intermediate step leading to the melting reaction. We believe this discovery of the hidden dormant phase, which explains the glass transition, will stimulate further structural investigations of disordered materials and advance fundamental knowledge on structure-forming kinetics.

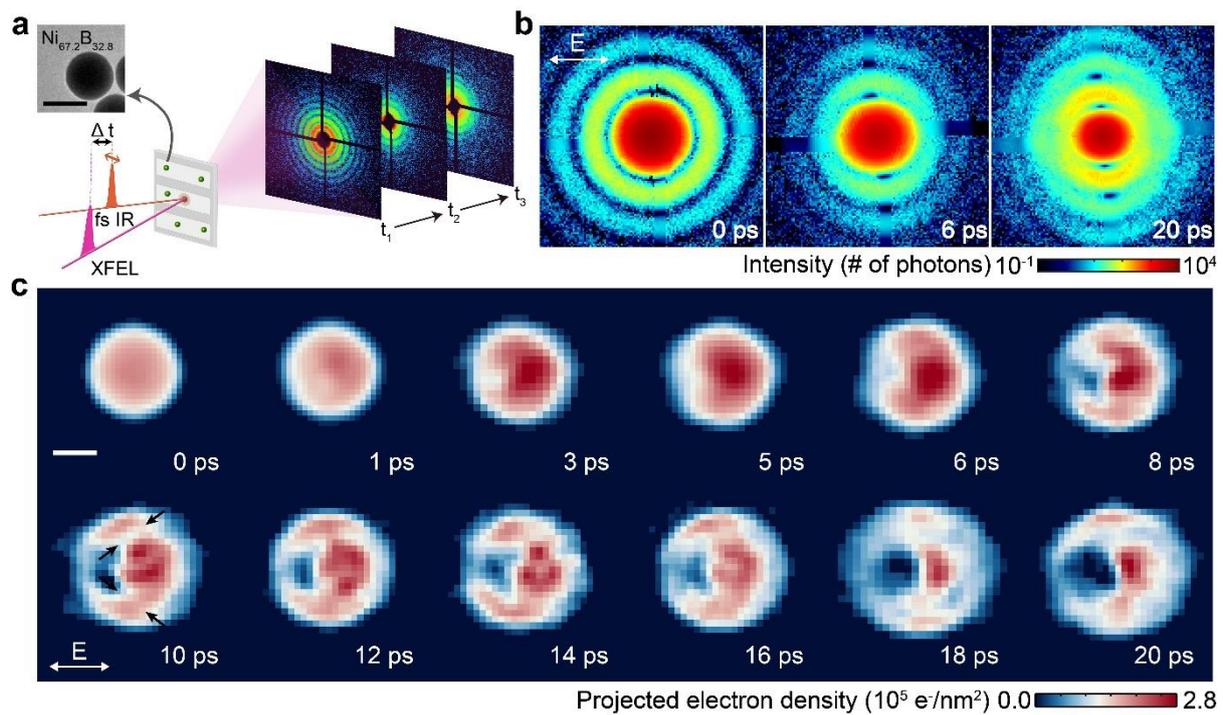

**Fig. 1: Direct observation of photoinduced glass-to-liquid transition in $Ni_xB_{1-x}$ MG NPs.** **(a)** Time-resolved single-shot imaging experiment via fs IR laser pumping and XFEL probing. **(b)** XFEL single-pulse diffraction patterns collected at increasing delay times indicate structural deformations in the $Ni_xB_{1-x}$ NPs induced by fs-IR laser. A dominant distortion from the circular Airy pattern is observed along the laser polarization direction. **(c)** Reconstructed images reveal photoinduced glass-to-liquid transition process. Local regions with reduced density are formed initially, which are then connected to form low-density bands (arrows in 10 ps indicating the low-density band, for instance). Disintegration of the $Ni_xB_{1-x}$ NPs is accelerated by lateral extension of channeled low-density structures in the later stage. The scale bar is 50 nm.

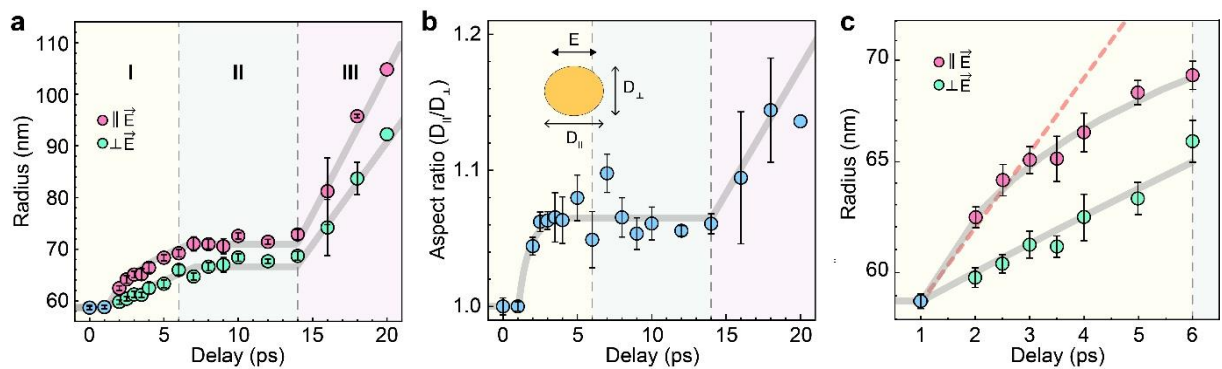

**Fig. 2: Radial expansion with three-stage reactions**. **(a)** Radial expansion of $Ni_xB_{1-x}$ NP shows three distinct stages: stage I with slow expansion (~ 1.2 km s$^{-1}$, until 6 ps), dormant stage II without expansion (between 6 and 14 ps), and rapidly expanding stage III (~ 4 km s$^{-1}$, after 14 ps). The expansion rate displays polarization-dependent anisotropic behavior, with faster expansion along the direction parallel to the laser polarization (magenta color). **(b)** The NP becomes anisotropic with a polarization-dependent expansion rate. **(c)** Upon entering stage II, the size expansion parallel to the laser polarization is parabolic, which is in contrast to the linear expansion observed for the direction perpendicular to the laser polarization.

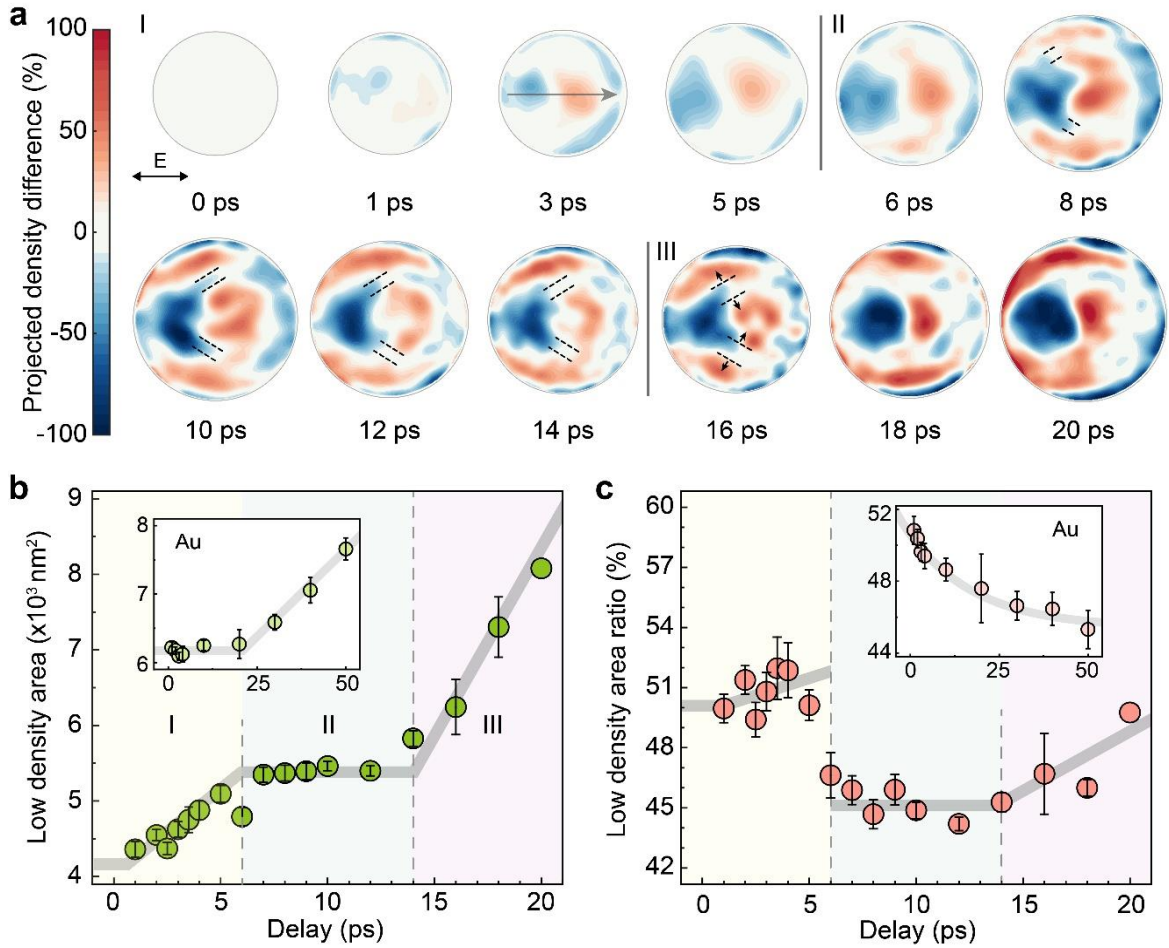

**Fig. 3: Ultrafast kinetics of low and high-density regions displaying three distinct development stages. (a)** Difference density maps show the evolution of low-density (blue) and high-density (red) regions. Localized low- and high-density regions are formed along the laser polarization direction in stage-I. The formation of low- and high-density regions creates dipolar shear forces parallel to the laser polarization direction. The gray arrow (drawn at 3 ps) shows the shear force direction. In stage II, channeling occurs as nearby localized low-density regions merge to form low-density bands (dashed black lines). Lateral expansion of low-density bands gradually increases the low-density area in stage III (black arrows indicate the expansion of low-density bands). **(b)** Total low-density area ($A_{low}$) is calculated. Between the gradually increasing areas in stages I and III, the low-density area in stage II remains unchanged. This dormant behavior contrasts with the melting in a typical metal, such as Au, which shows a monotonic increase (inset). **(c)** The ratio of low-density area to total area ($A_{low}/A_{total}$) displays a distinct pattern for each development stage. The fraction of low-density area is compared with Au NPs, which exhibit a steady decrease (inset).

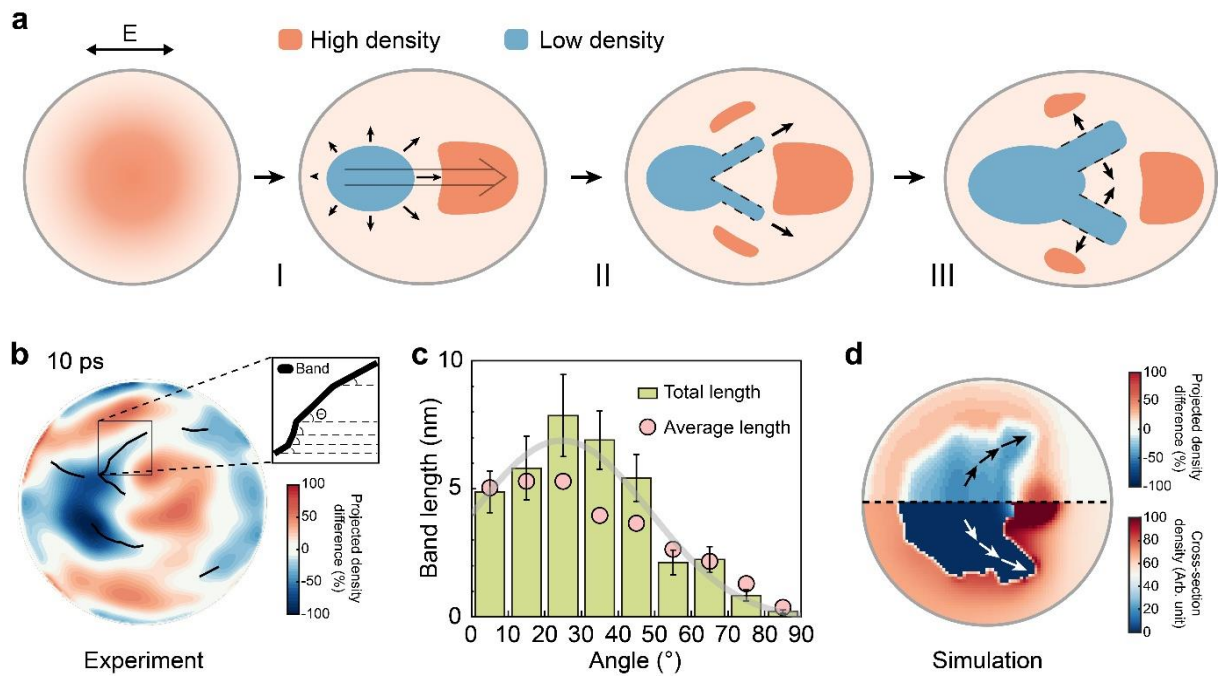

**Fig. 4: Dynamic construction of low-density bands describing the photoinduced glass-to-liquid transition.** (a) Schematics showing the microscopic evolution of low-density bands (blue) capture the photoinduced glass-to-liquid transition process. The homogeneous density distribution in intact MG NPs is disrupted by fs-IR laser-induced electronic photoexcitation, forming localized low- and high-density regions, leading to dipolar shear forces parallel to the laser field in stage I (gray arrow).The low-density region extends and penetrates the high-density region, forming bands in stage II. Lateral expansion of the low-density bands leads to rapid expansion and disintegration of the NP in stage III. Black arrows indicate the direction of low-density region expansion. (b) The low-density band is skeletonized to quantify the angular distribution. A difference density map from 10 ps is chosen to display the skeletonized band (thick black lines). (c) Low-density bands are formed with a directional preference between 20 and 40 degrees relative to the laser field. (d) Density variations obtained from MC simulations highlight the formation of distinct low-density and high-density regions during the band-forming, dormant phase (stage II). The upper half of the image shows a 3D projection

obtained by symmetric angular average of the cross-sectional density map in the lower half. The low-density regions show a directional preference along the diagonal relative to the laser field, consistent with experimental observations (black and white lines).

**Methods**

**Synthesis of spherical Ni$_x$B$_{1-x}$ MG NPs**

The Ni$_{67.2}$B$_{32.8}$ NPs with a diameter of 110 nm were synthesized via colloidal synthesis following previously established protocols[36,37]. Specifically, 10 mL of Milli-Q water was used as the solvent in a 25-mL beaker, and the stir bar speed was adjusted to create a visible vortex. First, 1 mM of nickel nitrate hexahydrate (2.91 mg) (Sigma-Aldrich) was added. The water solution of nickel nitrate hexahydrate was measured with a pipette for accuracy. Next, 28.84 mg of sodium dodecyl sulfate (SDS) (Sigma-Aldrich) was added. After the SDS had dissolved completely, 0.32 μL of oleic acid (Sigma-Aldrich), which had been previously dissolved in 100 μL of methanol (Sigma-Aldrich), was introduced. Finally, 3 mg of sodium borohydride (Sigma-Aldrich) was added to reduce the nickel ions, forming Ni$_x$B$_{1-x}$ NPs. The solution turned black immediately after the addition of sodium borohydride, indicating the formation of NPs. After the solution turned black, the beaker was removed from the stir plate, the stir bar was removed, and the solution was left undisturbed for at least 1 h to ensure reaction completion. The NPs were then centrifuged and sonicated in water multiple times to ensure thorough cleaning.

Based on the ICP results, the ratios of Ni and B within a single NP were 67.2% and 32.8%, respectively. Inductively coupled plasma optical emission spectrometry (ICP-OES) was performed using the Thermo Scientific ICAP 6300 Duo View Spectrometer to determine the composition of the Ni$_x$B$_{1-x}$ NPs. The NPs were digested in a 3% v/v nitric acid solution. Separate 100 ppm boron and nickel standards (Inorganic Ventures) were used. Synthesized NPs were drop-cast onto TEM grids with carbon film for TEM inspection. The bright-field and selected area electron diffraction (SAED) images were taken using the FEI Tecnai G2 F20 X-TWIN at 200 kV or the FEI Titan Environmental TEM at 300 kV (Fig. S1).

**Single-pulse time-resolved X-ray diffraction imaging**

Single-pulse, single-particle X-ray imaging experiments were conducted using a fixed X-ray energy of 9 keV at the PAL-XFEL. A fs Ti-sapphire laser with a wavelength of 800 nm and a pulse duration of 40 fs served as the pump source. The laser pulses were focused to a size of 174-μm root mean square (RMS) at the sample position. A linearly polarized laser was used to initiate localized deformation along the laser polarization direction in the $Ni_xB_{1-x}$ NPs. In this study, fs X-ray laser pulses were micron-focused using a pair of Kirkpatrick-Baez (K-B) mirrors, achieving a focal size of 2.6 μm × 2.9 μm and a focal length of 5 m. The spatial overlap of the specimens, the 174-μm focused IR laser, and the micron-focused XFEL was visually confirmed using an in-line microscope. The time zero of the interaction spot, where all three components (specimens, IR laser, and XFEL) were aligned, was verified by monitoring the absorption profile of a thin gallium nitride (GaN) crystal mounted at the interaction spot. The temporal resolution achieved with the PAL-XFEL was better than 0.5 ps without the use of a timing tool.

**Single-particle sample loading and single-pulse data acquisition**

Spherical single particles of $Ni_xB_{1-x}$ MG NPs, with a monodisperse and homogeneous sample morphology of 110 nm in diameter, were mounted on thin SiRN membranes. The membranes were custom-designed (Silson Ltd.) with multiple arrays on a frame. The experiment ensured that each NP was exposed to the IR laser and XFEL pulses only once. The shot-to-shot distance was set to 400 μm to ensure that a separation significantly larger than the laser footprint (<180 μm) prevents any overlap of irradiation zones or unintended exposure of adjacent NPs. This specialized protocol ensured accurate and isolated irradiation of the NPs. Single-pulse diffraction patterns were recorded using the JUNGFRAU detector, with a pixel size of 75 μm × 75 μm. The total detector area used was 2048 × 2048 pixels.

**Image reconstruction from phase retrieval of the coherent diffraction patterns**

Coherent X-ray diffraction patterns from single Ni$_x$B$_{1-x}$ NPs were obtained by exposing fresh NPs to single pulses of an fs XFEL. Each single-pulse diffraction pattern was phase retrieved to acquire an image of each independent NP[20]. This phase retrieval for acquiring a real image is not a fitting process but an optimization approach to the inverse problem through numerical iterations using a phasing algorithm[21]. Coherent diffraction patterns were measured with a sampling interval finer than the Nyquist frequency, which was determined by the inverse sample size used for the amplitude modulus of the Fourier transformation of the specimen.

Image reconstructions from acquired coherent diffraction patterns were performed using iterative phase retrievals with the generalized proximal smoothing algorithm. To enhance the signal-to-noise ratio (SNR), a 3 × 3-pixel summation was applied to the diffraction patterns. Each iterative process consisted of 2000 iterations, generating 200 independent, random phases. To further accelerate convergence, a guided image approach was implemented. In the initial (0$^{th}$) generation, reconstructions were performed using a random phase with diffraction pattern data. For subsequent (N$^{th}$) generations, the guide images were employed, where the guided image for each iteration was defined as: $\rho^{g,m+1} = \sqrt{\rho^{g,temp}(r) * \rho^{g,m}(r)}$. Here, $\rho^{g,temp}(r)$ represents the template image, and $\rho^{g,m}(r)$ is the image reconstructed from the previous generation. Generation was typically carried out 2 to 5 times for each image. The five best results were averaged to represent the final image, determined by the lowest error between the measured diffraction amplitude and the retrieved amplitude. The error metric used was: $R_{err} = \sum_k \left||I_{PR}(k)| - I_{exp}(k)\right|/|I_{exp}(k)|$. Here, $|I_{PR}(k)|$ is the Fourier amplitude of the phase retrieval result, and $|I_{exp}(k)|$ is the experimentally measured diffraction intensity. The summation excluded pixels with dead space in the JUNGFRAU detector. Data collection was repeated for each delay time, and multiple images were obtained for the same delay time (Supplementary Video 1). The image that best represented the average behavior of the images for the same delay time was selected.

**Data Availability**

The data supporting the findings of this study are available from the corresponding authors upon request.

**Acknowledgments**

The single-pulse time-resolved imaging experiments at PAL-XFEL were approved by the Korean Synchrotron Users Association (KOSUA). This work was supported by the National Research Foundation (NRF) of Korea (grant nos. 2022M3H4A1A04074153, RS-2024-00346711). MMW was supported by the National Science Foundation Graduate Research Fellowship under Grant No. DGE-1656518. MMW and XWG were supported by the grant DE-SC0021075 funded by the U.S. Department of Energy, Office of Science. MMW would like to thank Dr. Chaolumen Wu and Dr. Mehrdad Kiani for their guidance in synthesizing NPs.

**Author Contributions**

C.S. conceived and supervised the project. E.P. analyzed the experimental data overall. S.K. performed the MC simulations and image reconstructions. All authors contributed to the experiment. X.W.G. supervised the MG NP synthesis, and M.M.W. synthesized the MG NPs. E.P. and S.K. prepared the single-particle specimens on fixed targets. M.K., D.J., and I.E. operated the fs optical laser. C.S. and E.P. wrote the manuscript with input from all authors.

**Corresponding authors**

Correspondence can be addressed to C.S. (cysong@postech.ac.kr).

**Competing interests**

The authors declare no conflicts of interest.

**Additional Information**

Supplementary Information is available for this paper. Reprints and permissions information is available at www.nature.com/reprints.